\documentclass{article}
\usepackage[utf8]{inputenc}
\usepackage[margin=25mm]{geometry}
\usepackage{amsmath}
\usepackage{graphicx}
\usepackage{natbib}

\providecommand{\keywords}[1]
{
  \small	
  \textbf{\textit{Keywords---}} #1
}

\title{A Bridge between Cross-validation Bayes Factors and Geometric Intrinsic Bayes Factors}
\author{Yekun Wang$^{1}$, Luis Pericchi$^{1}$  \\
        \small $^{1}$University of Puerto Rico 
}
\date{} % Comment this line to show today's date

\begin{document}

\maketitle

\begin{abstract}
Model Selections in Bayesian Statistics are primarily made with statistics known as Bayes Factors, which are directly related to Posterior Probabilities of models. Bayes Factors require a careful assessment of prior distributions as in the Intrinsic Priors of \cite{berger1996intrinsic} and integration over the parameter space, which may be highly dimensional. Recently researchers have been proposing alternatives to Bayes Factors that require neither integration nor specification of priors. These developments are still in a very early stage and are known as Prior-free Bayes Factors, Cross-Validation Bayes Factors (CVBF), and Bayesian "Stacking." This kind of method and Intrinsic Bayes Factor (IBF) both avoid the specification of prior. However, this Prior-free Bayes factor might need a careful choice of a training sample size. In this article, a way of choosing training sample sizes for the Prior-free Bayes factor based on Geometric Intrinsic Bayes Factors (GIBFs) is proposed and studied. We present essential examples with a different number of parameters and study the statistical behavior both numerically and theoretically to explain the ideas for choosing a feasible training sample size for Prior-free Bayes Factors. We put forward the "Bridge Rule" as an assignment of a training sample size for CVBF's that makes them close to Geometric IBFs.  We conclude that even though tractable Geometric IBFs are preferable, CVBF's, using the Bridge Rule, are useful and economical approximations to Bayes Factors.

\end{abstract}\hspace{30pt}


\section{Background}

\subsection{Cross-validation Bayes Factors}

Cross-validation Bayes Factor proposed by \cite{hart2019prior} is a direct way to apply Cross-validation to Bayes factors. Assume that ${ X }_{ 1 },...,{ X }_{ n }$ are independent and identically distributed variables from density $f$. Let $\{ f(\cdot |\theta ):\theta \in \Phi \}$ and $\{ g(\cdot |\lambda ):\lambda \in \Lambda \}$be parametric models for $ f,$ where $\Phi $ and $\Lambda$ belong to some different (or the same) dimensional Euclidean spaces. Hence, the likelihood functions are ${ L }_{ 1 }(\theta )=\prod _{ i=1 }^{ n }{ f({ X }_{ i }|\theta ) }$ and ${ L }_{ 2 }(\lambda )=\prod _{ i=1 }^{ n }{ g({ X }_{ i }|\lambda ) }.$ The first step to compute Bayes factor is to split the data matrix into two disjiont parts, which initially and for convenience we take to be \[{ { X }_{ T } }^{ (l) }=({ { X }_{ 1 } }^{ (l) },...,{ { X }_{ m } }^{ (l) })\] and \[{ { X }_{ V } }^{ (l) }=({ { X }_{ m+1 } }^{ (l) },...,{ { X }_{ n } }^{ (l) }),\] and $l$ refers to the particular data split, where $l=1,...,L$, usually $L=\binom{n}{m}.$ These two subsets of the data are training set and validation set in cross-validation. Now, let ${ { \hat { \theta  }  }_{ m } }^{ (l) }$ and ${{ \hat { \lambda  }  }_{ m }}^{(l)}$ be the maximum likelihood estimators of $\theta$ and $\lambda,$ respectively, that are computed from the data set ${{ X }_{ T }}^{(l)}.$ At the last step, we evaluate the likelihood functions using validation set, that is, ${{ X }_{ V }}^{(l)}$. At this point $f(\cdot |{{ \hat { \theta  }  }_{ m }}^{(l)})$ and $g(\cdot |{{ \hat { \lambda  }  }_{ m }}^{(l)})$ are two simple models for the underlying distribution of ${{ X }_{ i }}^{(l)},$ and therefore we have Cross-validation Bayes factor \[ B({ { X }_{ T } }^{ (l) },{ { X }_{ V } }^{ (l) })=\frac { \prod _{ i=m+1 }^{ n }{ f({ { X }_{ i } }^{ (l) }|{ { \hat { \theta  }  }_{ m } }^{ (l) }) }  }{ \prod _{ i=m+1 }^{ n }{ g({ { X }_{ i } }^{ (l) }|{ { \hat { \lambda  }  }_{ m } }^{ (l) }) }  } . \] 

However, such a cross validation statistics, depends on the particular training sample employed. If we take a geometric mean of $L$ repeats of , then it is no longer dependent on the particular training sample. CVBF becomes \[ AvgB({ X }_{ T },{ X }_{ V })={ [\prod _{ l=1 }^{ L }{ \frac { \prod _{ i=m+1 }^{ n }{ f({ { X }_{ i } }^{ (l) }|{ { \hat { \theta  }  }_{ m } }^{ (l) }) }  }{ \prod _{ i=m+1 }^{ n }{ g({ { X }_{ i } }^{ (l) }|{ { \hat { \lambda  }  }_{ m } }^{ (l) }) }  }  } ] }^{ 1/L },\] where $L$ used above represents $\binom{n}{m}$.

For example, if there is a model that follows a normal distribution with an unknown mean and a known variance. Denote these two parameters by $\, \mu \,$ and $\, \sigma,  \,$ respectively. The Maximum likelihood estimator for the mean is just the sample mean. Computing the estimator using the training set and evaluating likelihood functions using the validation set, we take the ratio of likelihood functions of models we compared. Finally, the geometric average over all the possible training samples of size $m$ is calculated as \[AvgB({ X }_{ T },{ X }_{ V })=exp\{ \frac { 1 }{ L } \sum _{ l=1 }^{ L }{ logB({ { X }_{ T } }^{ (l) },{ { X }_{ V } }^{ (l) }) } \} .\]

\subsection{Geometric Intrinsic Bayes Factors}

For Intrinsic Bayes Factors, we calculate a posterior using training samples on the prior distribution, and then evaluate the marginal likelihood functions on both models using the validation set. If we take a geometric mean of the IBFs, then it becomes a Geometric Intrinsic Bayes Factors, which is expressed as below (Here we use same notations as section 1.1),
\[\\ { B }({ X_{ V } }^{ (l) }|{ X_{ T } }^{ (l) })={ [\prod _{ l=1 }^{ L }{ \frac { \int { f({ X_{ V } }^{ (l) }|\theta ) } { \pi  }(\theta |{ X_{ T } }^{ (l) })d\theta  }{ \int { g({ X_{ V } }^{ (l) }|\lambda ) } \pi (\lambda |{ X_{ T } }^{ (l) })d\lambda  }  } ] }^{ 1/L },\]
 where $L$ used above represents $\binom{n}{m}$.

\subsection{Corrected Intrinsic Bayes Factors}

If a prior is a proper prior, it is supposed to integrate to 1. For example, a normal distribution is integrating to one while a uniform distribution is not integrating to one for whatever the choice of the constant c is because $\int _{ -\infty  }^{ \infty  }{ cdx } =\infty.$ Therefore, distributions that cannot integrate to 1, such as, a uniform distribution, are called improper priors, which are often considered as uninformative priors. Uninformative priors are sensible in Hypothesis Testing problems since they should take into account the null considered.  The priors of arithmetic intrinsic Bayes Factors integrate to 1 absolutely. However, geometric intrinsic priors usually integrate to a finite positive constant $c>0$ so they need a correction $1/c$.  As \cite{berger1996intrinsic} proposed, the geometric intrinsic prior is \[{ { \pi  } }^{ GI }(\theta )={ { \pi  } }^{ N }(\theta )exp\{ \int { \cdots \int { log{ { BF }_{ 0,1 } }^{ N }({ X }_{ T }) } f({ X }_{ T }|\theta )d{ X }_{ 1 }\cdots d{ X }_{ m } } \},  \] where ${ { BF }_{ 0,1 } }^{ N }({ X }_{ T })=\frac { f({ X }_{ v }|{ \theta  }_{ 0 }) }{ \int { f({ X }_{ v }|\theta )\pi (\theta |{ X }_{ T })d\theta  }  } $, $X_{1},..., X_{m}$ are in the set $X_{T}$, $\theta$ is the parameter under analysis, $\theta_{0}$ is a constant value for $\theta$ under the simpler model and $N$ stands for non-informative.

And the integration for Geometric prior is \[\int { { { \pi  } }^{ GI }(\theta ) } d\theta =c\neq 1,\] where $c$ is a positive constant.

Therefore, the Corrected Geometric prior is \[ { { { \pi  } }^{ cGI } }(\theta )=\frac { { { \pi  } }^{ GI }(\theta ) }{ c }, \] which will be integrating to one. 

After the corrections for Geometric Prior, then we need to modify the GIBF so that it becomes a Corrected GIBF.

\section{Problem Statement}

For the popular method, Intrinsic Bayes Factor, it only requires the number of parameters as a training sample size, which is a small number for an extensive collection of problems. However, it cannot avoid integration over the parameter space, even if it has an intrinsic prior distribution. 

On the other hand, Cross-validation Bayes Factor seems to be quite simple because it does not have to choose a prior distribution and does not require integration. In this regard, if we can adapt CVBF to be a Bayesian method and establish a bridge between CVBF and IBF (Actually, it should be GIBF), then we could find a double cure. The first is to circumvent the computational difficulties related to GIBF, and the second is to make CVBF truly Bayesian statistics. 

In a word, with the help of GIBF, CVBF may become a useful approximation if one finds a hidden prior distribution and a reasonable training sample size. 

Another crucial thing is that, can we use CVBF in the Model Selections without other concerns? The stability and consistency are also considered in this article.

\section{Normal Means Problem}
Let us begin with the simplest problem to gain insight into the interplay between GIBF and CVBF. We are here analyzing a hypothesis testing with a null hypothesis ${ H }_{ 0 }$, in which is a normal distribution with mean $\theta=\theta_{0}$ and variance ${ \sigma  }^{ 2 },$ where ${ \sigma  }^{ 2 }$ is known. While the alternative hypothesis ${ H }_{ 1 }$ is a normal distribution with mean different from $\theta_{0}$ and the same variance ${ \sigma  }^{ 2 }$.

\subsection{CVBF in normal means problem}
We apply the expression in section 1.1, CVBF becomes \[ B({ { X }_{ T } }^{ (l) },{ { X }_{ V } }^{ (l) })\, =\, exp\{ -\frac { (n-m) }{ 2{ \sigma  }^{ 2 } } [{ { ({ { { \bar { { x } }  }_{ m } }^{ (l) } } }-{ { \bar { { x } }  }_{ n-m } }^{ (l) }) }^{ 2 }-{ { ({ { { \bar { { x } }  }_{ n-m } }^{ (l) } } }-{ { \theta  }_{ 0 } }) }^{ 2 }]\} ,  \] where ${ \bar { { x } } _{ m } }^{ (l) }$ is the mean of the training samples while ${{ \bar { { x } }  }_{ n-m }}^{(l)}$ is the mean of the validation set.

In general, CVBF can be expressed as 
\begin{equation}
{ { BF }_{ 1,0 } }^{ CV }={ e }^{ -\frac { 1 }{ 2 } (\frac { n }{ m } { Z }^{ 2 }-{ \tilde { Z }  }^{ 2 }) },
\end{equation}
where $\, Z^{2} \,$ stands for a Chi-squared distribution with 1 degree of freedom, $\, { {\tilde { Z }}^{2}  }\,$ is a non-central Chi-squared distribution with 1 degree of freedom and non-centrality parameter $\frac { (n-m){ (\theta -{ \theta  }_{ 0 }) }^{ 2 } }{ { \sigma  }^{ 2 } }. $

The geometric average of CVBFs is \[{ Avg{ B }F^{ CV } }_{ 1,0 }=exp\{ -\frac { n-m }{ 2K{ \sigma  }^{ 2 } } [(\frac { 1 }{ m } +\frac { 1 }{ n-m } ){ \sigma  }^{ 2 }{ \chi  }^{ 2 }(K)-\frac { { \sigma  }^{ 2 } }{ n-m } { \tilde { \chi  }  }^{ 2 }]\}, \]where $K$ is the number of simulation, in this case, $K=\binom{n}{m}$, which includes all the possibilities of training sample sets, and ${ \chi  }^{ 2 }(K)$ is a Chi-square distribution with $K$ degrees of freedom and ${ \tilde { \chi  }  }^{ 2 }$ is a non-central Chi-square distribution with K degrees of freedom and non-centrality $\frac { K{ (\theta -{ \theta  }_{ 0 }) }^{ 2 } }{ { \sigma  }^{ 2 }/(n-m) } $.

Hence, under the alternative model, we have \[ E(log{ { BF }_{ 1,0 } }^{ CV }|{ M }_{ 1 })=E(log{ { AvgBF }_{ 1,0 } }^{ CV }|{ M }_{ 1 })=-\frac { 1 }{ 2 } (\frac { n }{ m } -1-\frac { (n-m){ (\theta -{ \theta  }_{ 0 }) }^{ 2 } }{ { \sigma  }^{ 2 } } ),\] and the variance of logarithm of CVBF is \[ Var(log{ { BF }_{ 1,0 } }^{ CV }|{ M }_{ 1 })=\frac { { n }^{ 2 } }{ 2{ m }^{ 2 } } +\frac { 1 }{ 2 } +\frac { (n-m){ (\theta -{ \theta  }_{ 0 }) }^{ 2 } }{ { \sigma  }^{ 2 } }.   \]

\subsection{Corrected GIBF in normal means problem}

We also apply the formula in section 1.2 to express GIBF in normal means problem. For simplicity, we use a uniform distribution as a non-informative prior distribution, we denote ${ { \pi  }^{ N } }(\theta )=1$. IBF has been expressed by \[{ B }_{ 1,0 }({{ X }_{ V }}^{(l)}|{{ X }_{ T }}^{(l)})=\sqrt { \frac { m }{ n }  } exp\{ -\frac { 1 }{ 2{ \sigma  }^{ 2 } } [m{ ({\bar { { x } }_{ m }}^{(l)} -{ \theta  }_{ 0 }) }^{ 2 }-n{ ({\bar { { x } }_{ n }}^{(l)} -{ \theta  }_{ 0 }) }^{ 2 }]\}, \] where ${{ \bar { { x } }_{ m }}}^{(l)}$ is the mean of training samples and ${ \bar { { x } }_{ n } }^{(l)}$ is the mean of the data for split $l$.

Training sample size 1 is a minimal size because it coincides with the number of parameters in the model. The corrected constant for GIBF we computed is $\sqrt { e }$. After this correction, we have corrected IBF \[{ { B }_{ 1,0 } }^{ cI }({ X }_{ V }|{ X }_{ T })=\sqrt { \frac { e }{ n }  } exp\{ -\frac { 1 }{ 2{ \sigma  }^{ 2 } } [{ ({ x }_{ i }-{ \theta  }_{ 0 }) }^{ 2 }-n{ (\bar { { x } } _{ n }-{ \theta  }_{ 0 }) }^{ 2 }]\},  \] where $i=0,1,...,n$.

Therefore, corrected GIBF can be expressed by \[{ { B }_{ 1,0 } }^{ cGI }({ X }_{ V }|{ X }_{ T })=\sqrt { \frac { e }{ n }  } exp\{ -\frac { 1 }{ 2{ \sigma  }^{ 2 } } [\frac { \sum _{ i=1 }^{ n }{ { ({ x }_{ i }-{ \theta  }_{ 0 }) }^{ 2 } }  }{ n } -n{ (\bar { { x } } _{ n }-{ \theta  }_{ 0 }) }^{ 2 }]\}.  \]

Then, under model 1, the expectation of logarithm of corrected IBF is \[ E(log{ { BF }_{ 1,0 } }^{ cI }|{ M }_{ 1 })=E(log{ { BF }_{ 1,0 } }^{ cGI }|{ M }_{ 1 })=\frac { 1 }{ 2 } log\frac { e }{ n } +\frac { 1 }{ 2 } \frac { (n-1){ (\theta -{ \theta  }_{ 0 }) }^{ 2 } }{ { \sigma  }^{ 2 } } ,\] and the variance is \[ Var(log{ { BF }_{ 1,0 } }^{ I }|{ M }_{ 1 })=1-\frac { 1 }{ n } +\frac { (n-1){ (\theta -{ \theta  }_{ 0 }) }^{ 2 } }{ { \sigma  }^{ 2 } } . \]

We denote here the training sample size as $m$, and yet a GIBF with a minimal training sample size performs quite well in most scenarios, in which some of these will be presented in the following section.

\subsection{Bridge Rule and consistency analysis}
When the null hypothesis is correct, to let GIBF and CVBF be approximately equivalent, a training sample size of CVBF should be assigned. We propose that a training sample size for CVBF should be \[{ M }_{ cv }=\frac { N }{ logN }, \] where $N$ is the sample size. Under this Rule, we pass the GIBF consistency under the null model to CVBF.  It is worth mentioning that the Bridge Rule $\frac{N}{logN}$, at least at the domain $\{N|N\in (0,500],N\in Z\}$, can be approximated by linear regression. In the Figure 1, $\frac{N}{logN}$ is fitted by a linear equation $y = 6.2622+0.1519N$. Therefore, the bridge rule is approximately a straight line with a slope of 0.152.

\begin{figure}
    \centering
    \includegraphics[scale=0.5]{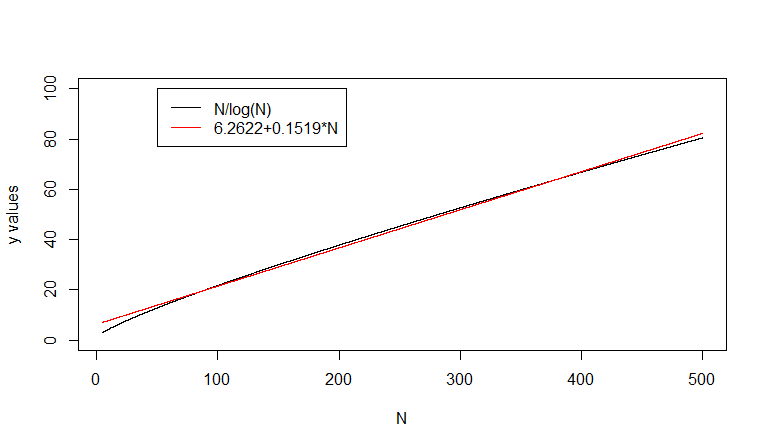}
    \caption{Linear Regression}
    \label{Figure 1}
    \medskip
    \small
    The black line is our bridge rule at the domain $(0,500]$, while the red line highlights a linear equation which approximates the bridge rule function.
\end{figure}

On the other hand, this raises a problem: Do we obtain consistency for CVBF and GIBF under the alternative? 
 
By equation $(1)$, under the rule, the rate for CVBF could be changed. Under $H_{0}$,
\[log{ { BF }_{ 1,0 } }^{ CV }=-\frac { 1 }{ 2 } (log(n)-1){ Z }^{ 2 },\] where ${Z}^{2}$ has a standard Chi-square distribution. \[P(-\frac { 1 }{ 2 } (log(n)-1){ Z }^{ 2 }<-{ C }_{ n }|H_{0})\rightarrow 1\] at a rate $log(n)$, where $C_{n}$ is a positive constant. 

When it comes to corrected GIBF,
\[P(log{ { BF }_{ 1,0 } }^{ CGI }<{ -C }_{ n }|{ H }_{ 0 })\rightarrow 1\] at a rate $log(n)$.

On the other hand, we analyze the scenario under $H_{1}$,
\[log{ { BF }_{ 1,0 } }^{ CV }=-\frac { n\frac { logn-1 }{ logn }  }{ 2{ \sigma  }^{ 2 } } [{ ( { \bar{ x }_{ m } } -\theta ) }^{ 2 }-{ (\theta -{ \theta  }_{ 0 }) }^{ 2 }].\]
\[P(-\frac { n\frac { logn-1 }{ logn }  }{ 2{ \sigma  }^{ 2 } } [{ ( { \bar{ x }_{ m } } -\theta ) }^{ 2 }-{ (\theta -{ \theta  }_{ 0 }) }^{ 2 }]>{ C }_{ n }|{ H }_{ 1 })\rightarrow 1\]  at a rate n.

In contrast, 
\[P(\frac { 1 }{ 2 } -\frac { 1 }{ 2 } logn-\frac { 1 }{ 2{ \sigma  }^{ 2 } } [\frac { \sum { { ({ x }_{ i }-{ \theta  }_{ 0 }) }^{ 2 } }  }{ n } -n{ (\bar { x } -{ \theta  }_{ 0 }) }^{ 2 }]>{ C }_{ n }|{ H }_{ 1 })\rightarrow 1.\] Since $n{ (\bar { x } -{ \theta  }_{ 0 }) }^{ 2 }$ term dominates the inequality, $P(log{ { BF }_{ 1,0 } }^{ CGI }>{ C }_{ n }|{ H }_{ 1 })\rightarrow 1$ at a rate n. 

If we ignore the constant terms, such as $\frac { { (\theta -{ \theta  }_{ 0 }) }^{ 2 } }{ { \sigma  }^{ 2 } }$, we can simply express the equation for expectations of CVBF and corrected IBF as below, \[E(log{ { BF }_{ 1,0 } }^{ CV }|{ M }_{ 1 })\rightarrow -log(n)+n(1-\frac { 1 }{ log(n) } ),\] \[E(log{ { BF }_{ 1,0 } }^{ cI }|{ M }_{ 1 })\rightarrow -log(n)+n,\] respectively, as n goes to infinity. It is easy to see that $E(log{ { BF }_{ 1,0 } }^{ CV }|{ M }_{ 1 })\rightarrow \infty $ at $O(n)$ while $E(log{ { BF }_{ 1,0 } }^{ cI }|{ M }_{ 1 })\rightarrow \infty $ also at $O(n)$ when $n$ goes to infinity, which means $\frac { E(log{ { BF }_{ 1,0 } }^{ CV }|{ M }_{ 1 }) }{ E(log{ { BF }_{ 1,0 } }^{ cI }|{ M }_{ 1 }) } \rightarrow 1.$ Under the circumstance of choosing $n/log(n)$ as a training sample size of Cross-Validation Bayes Factors, CVBF and corrected Geometric Intrinsic Bayes Factors achieve consistency under both null hypothesis and alternative assumption.

After illustrating the above, we can conclude that after the Rule, GIBF and CVBF have consistency under both the null model and the alternative model. Furthermore, they tend to the correct model at the same rate of convergence, quite a promising result. 

It can be argued that the constant $c$ to correct the GIBF may be difficult to compute in complex problems. However, even if we do not calculate the constant $c$ exactly on each problem, but use the correction for Normal problems. Still, asymptotically both methods CVBF and GIBF are expected to be consistent at the same rate since the correction factor $c$ is just a fixed constant, bounded away both from zero and infinity.  

\subsection{Performances and simulations}

Under the null hypothesis, we generate the data of size 100 by a normal distribution of $N(0,1)$. On the contrary, under the alternative model, we generate the data of the same size by a normal distribution of $N(0.25,1)$. We analyze type I errors and type II errors under different training sample sizes (from 5 to 95 with spacing 5). At this point, we employ Receiver Operating Characteristic (ROC) to evaluate scores via the Area Under Curve (AUC).

\begin{figure}
    \centering
    \includegraphics[scale=0.5]{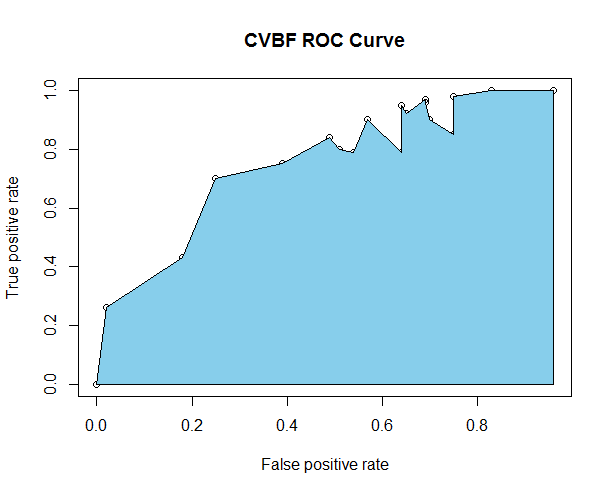}
    \caption{ROC Curve for CVBF}
    \label{Figure 2}
    \medskip
    \small
    The blue area is the area under curve (AUC) of CVBF, the thresholds here are training samples, which vary from 5 to 95 with spacing 5 when the sample size is 100.
\end{figure}

\begin{figure}
    \centering
    \includegraphics[scale=0.5]{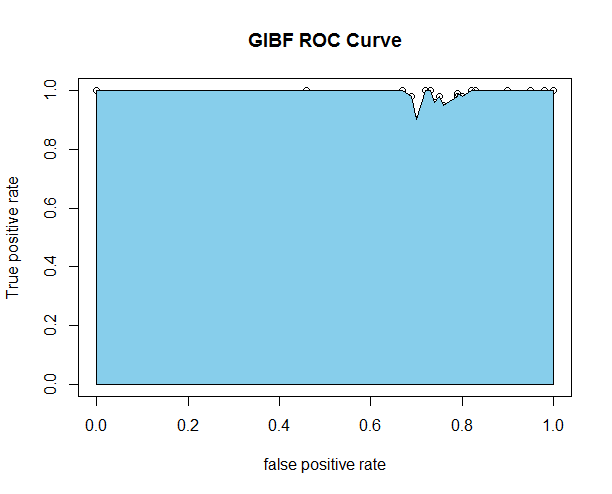}
    \caption{ROC curve for GIBF}
    \label{Figure 3}
    \medskip
    \small
    The blue area is the area under the curve (AUC) of GIBF, the thresholds here are training samples, which vary from 5 to 95 with spacing 5 when the sample size is 100.
\end{figure}

Figure 2 and Figure 3 suggest that the area under the curve of CVBF is 0.7125, while the AUC of GIBF is 0.9960. An area of 1 represents a perfect test, while an area of 0.5 represents a worthless test, which indicates that GIBF is an excellent and better method than CVBF, and one does not need to specify a training sample size. However, CVBF is still an attractive, simple method, although it needs a careful assessment of its training sets. Our proposal that we called a bridge rule seems to be sensible.

Now we vary the sample size but fix the training sample size for GIBF as 1. According to our Rule, the training sample size for CVBF would depend on the training sample size of GIBF. Under the null hypothesis, the data points are generated from a normal distribution of $N(0,1)$. In another scenario, under the alternative model, the data points are drawn from a normal distribution of $N(0.25,1)$. The sample size varies from 5 to 500, with spacing 5.

\begin{figure}
    \centering
     \includegraphics[scale=0.50]{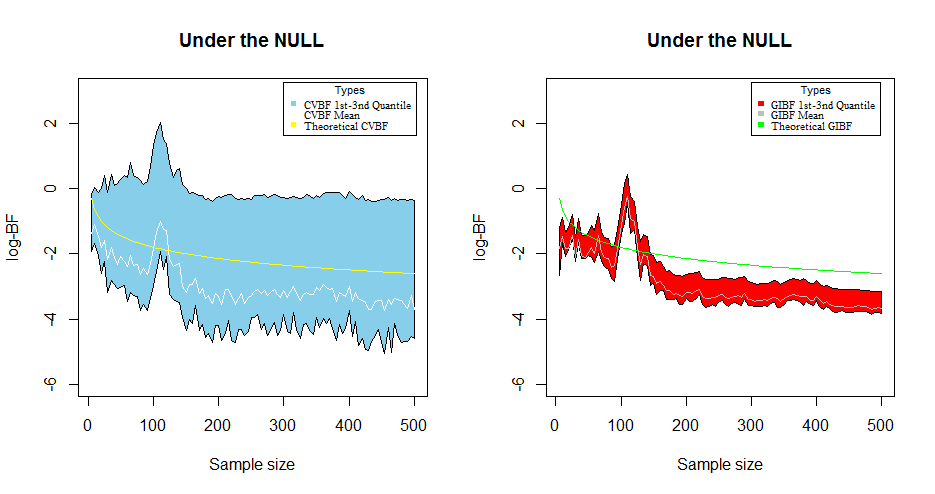}
    \caption{Consistency under the Null in One-parameter Normal Case}
    \label{Figure 4}
    \medskip
    \small
    In this figure, the red area is the range from the first quantile to the third quantile on 1000 simulations of the log of GIBF, while the blue area is the range from the first quantile to the third quantile on 1000 simulations of the log of CVBF. Moreover, the white line and the grey line are means on the 1000 simulations of the log of CVBF and GIBF, respectively. The yellow line in the left panel refers to the theoretical result of the expectation for CVBF. By contrast, the green line in the right panel is the expectation of GIBF. 
\end{figure}
\begin{figure}
    \centering
    \includegraphics[scale=0.50]{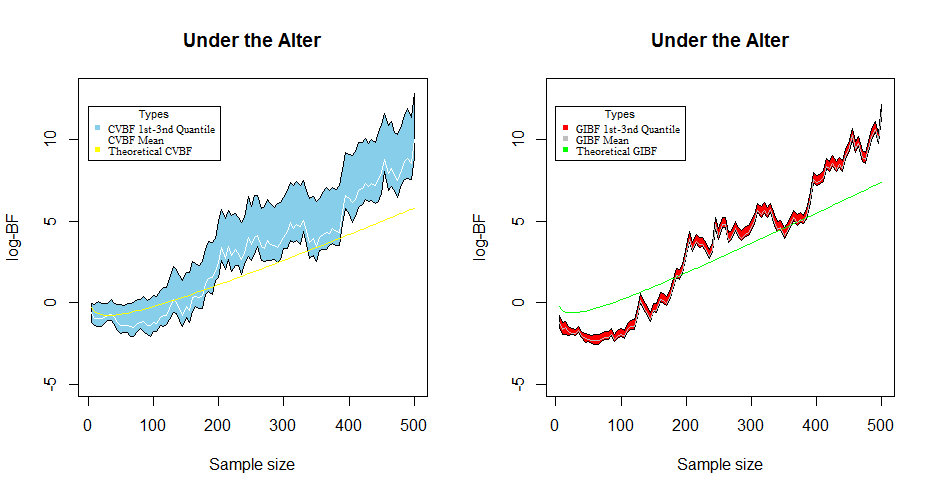}
    \caption{Consistency under the Alternative in One-parameter Normal Case}
    \label{Figure 5}
    \medskip
    \small
    In this figure which relates to the alternative hypothesis model, the red area is the range from the first quantile to the third quantile on 1000 simulations of the log of GIBF; while the blue area is the range from the first quantile to the third quantile on 1000 simulations of the log of CVBF. Moreover, the white line and the grey line are means on the 1000 simulations of the log of CVBF and GIBF, respectively. The yellow line refers to the theoretical result of the expectation for CVBF. By contrast, the green line is the expectation of GIBF. 
\end{figure}

From Figure 4 and Figure 5, we can observe that CVBF and GIBF are consistent under the null hypothesis, which happens under the alternative model. Furthermore, the expectations coincide with the simulations. We use GIBF as a guide for choosing the training sets for CVBF and take advantage of the simplicity of CVBF for computing Bayes Factors. The only shortcoming is that we sacrifice the variability. The variance of CVBF is larger than one in GIBF, which we can also conclude from Section 3.1 and Section 3.2 if we vary $n$, and take $m=1$, $\sigma=1$, $\theta_{0}=0$, and $\theta=0.25$ when we compute the expectations of variances. Fortunately, as \cite{tukey1963less} suggested, we can overcome this large variability by trimming the two ends of the ordered sequence of values of Bayes factors in our simulations, which reduces a significant width of the variances for CVBF.

\section{Exponential case}

The probability density function of an exponential is $f(x|\beta )=\beta exp\{ -\beta x\}$, where $\beta>0$. The null hypothesis is ${ H }_{ 0 }: \beta =\beta_{0},$ and the alternative is ${ H }_{ 1 }: \beta \neq \beta_{0}$. 

\subsection{IBF in exponential}

The prior for IBF here we use Jefferys prior ${ \pi  }^{ N }(\beta )=\frac { 1 }{ \beta  }$. Moreover, for simplicity, the training sample size of IBF equals one, the number of parameters.

Then the Intrinsic Bayes Factor is going to be
\[{ { BF }_{ 1,0 } }^{ I }=\frac { \Gamma (n){ x }_{ l }^{ (l) } }{ { (n\bar { x } ) }^{ n }{ { \beta  }_{ 0 } }^{ n-1 }exp\{ -{ \beta  }_{ 0 }(n-1){ \bar { { x } }  }_{ n-1 }^{ (l) }\}  }, \] where $x_{l}^{ (l) }$ is one data point, ${ \bar { { x } }_{n-1}}^{ (l) }$ is the mean of the rest $n-1$ data points and $\Gamma$ is a Gamma function.

The correction factor for GIBF is $exp\{\psi (1)\}$, where $\psi (1)$ is a digamma function at 1. Hence, the log of the corrected IBF is 

\[log{ { BF }_{ 1,0 } }^{ cI }= log(\Gamma (n))+log(\Gamma (1,\beta ))-nlog(n\Gamma (n,n\beta ))-(n-1)log({ \beta  }_{ 0 })+{ \beta  }_{ 0 }(n-1)\Gamma (n-1,(n-1)\beta )-\psi (1),\]
where $\Gamma (n)$ is a Gamma function, $\Gamma (1,\beta )$, $\Gamma (n,n\beta )$, $\Gamma (n-1,(n-1)\beta )$ are Gamma distributions.

After taking the expectations of each term, the expectation of the log of GIBF is
\[E[log({ { BF }_{ 1,0 } }^{ cGI })|{ M }_{ 1 }]=E[log({ { BF }_{ 1,0 } }^{ cI })|{ M }_{ 1 }]=log(\Gamma (n))-n\psi (n)+(n-1)log(\frac { \beta  }{ { \beta  }_{ 0 } } )+(n-1)\frac {{ \beta  }_{0}}{ { \beta  } }, \] where $\psi (n)$ is a digamma function at n.

\subsection{CVBF in exponential}

Cross-validation Bayes Factor in the exponential case is \[{ { B }_{ 1,0 } }^{ CV }={ (\frac { 1 }{ { \beta  }_{ 0 }{ \bar { x }  }_{ m }^{ (l) } } ) }^{ n-m }exp\{ -\frac { (n-m){ \bar { x }  }_{ n-m }^{ (l) } }{ { \bar { x }  }_{ m }^{ (l) } } +{ \beta  }_{ 0 }(n-m){ \bar { x }  }_{ n-m }^{ (l) }\}, \] the log of the CVBF is
\[log{ { BF }_{ 1,0 } }^{ CV }= -(n-m)log{ ({ \beta  }_{ 0 }\Gamma (m,m\beta )) }+{ \beta  }_{ 0 }(n-m)\Gamma (n-m,(n-m)\beta )-(n-m)\beta '(n-m,m,1,\frac { m }{ n-m } ),\]where $\Gamma$ is a Gamma distribution and $\beta '$ is a Beta Distribution of the Second Kind.

Then, we attain the expectation of the log of CVBF.
\[E(log{ { BF }_{ 1,0 } }^{ CV }|{ M }_{ 1 })=(n-m)log(\frac { \beta  }{ { \beta  }_{ 0 } } )-(n-m)\psi (m)+(n-m)log(m)+\frac { { \beta  }_{ 0 } }{ \beta  } (n-m)-(n-m)\frac { m }{ m-1 }. \]

\subsection{Approximations and Bridge Rule}

We need to make some approximations. One of properties for digamma function is
\[\psi (n)= \sum _{ k=1 }^{ n-1 }{ \frac { 1 }{ k }  } -\gamma, \] where $\gamma$ is an Euler-Mascheroni constant, as in \cite{johnson1970continuous}, we have an approximation \[\psi (n)\approx log(n-\frac { 1 }{ 2 } ).\]

And also, for $log\Gamma (n)$,
\[log\Gamma (n)\approx -(n-1)+(n-1)log(n-1)+\frac { 1 }{ 2 } log(2\pi (n-1)).\]

After attaining approximations, we can figure out the Bridge Rule of training sets for CVBF in the exponential case. Without being surprised, under the null hypothesis, CVBF approximates to GIBF when using the bridge rule $\frac { N }{ logN }$ for a large sample size.

\subsection{Consistency}

Under the null hypothesis, the expectation of the log of IBF the log of CVBF are 
\[E(log{ BF }^{ CI }|{ M }_{ 0 })=(n-1)log(n-1)+\frac { 1 }{ 2 } log(2\pi (n-1))-nlog(n-\frac { 1 }{ 2 } ), \]and
\[E(log{ BF }^{ CV }|{ M }_{ 0 })=(n-\frac { n }{ log(n) } )[log(\frac { \frac { n }{ log(n) }  }{ \frac { n }{ log(n) } -\frac { 1 }{ 2 }  } )-\frac { 1 }{ \frac { n }{ log(n) } -1 } ], \] respectively. 

They both go to $-\infty$ at a rate of $log(n)$.

Under the alternative, the expectations of the log-IBF and log-CVBF both go to $\infty$ at a rate of $n$. Based on two situations above, we can conclude that CVBF and GIBF converge for a considerably big $n$ under Bridge Rule $\frac { N }{ logN }$. 

\subsection{Simulations in exponential}

To make a scenario, we suppose a hypothesis test, which is a null model $\beta=0.2$ against an alternative $\beta \neq 0.2$. Fixing the size of data as 100, we vary the $\beta$ to see the tendencies of CVBF and GIBF. We observed from Figure 6 that CVBF perfectly coincides with GIBF with tiny gaps under the Rule. Note that CVBF is just slightly sensitive to the parameter for the reason that, roughly, CVBF is in favor of the null hypothesis at the domain of $(0.1662,0.2608)$ while GIBF favors the null model at $(0.1636,0.2640)$. The interval for selecting the null from CVBF is slightly narrower than the one from GIBF.

\begin{figure}
    \centering
    \includegraphics[scale=0.5]{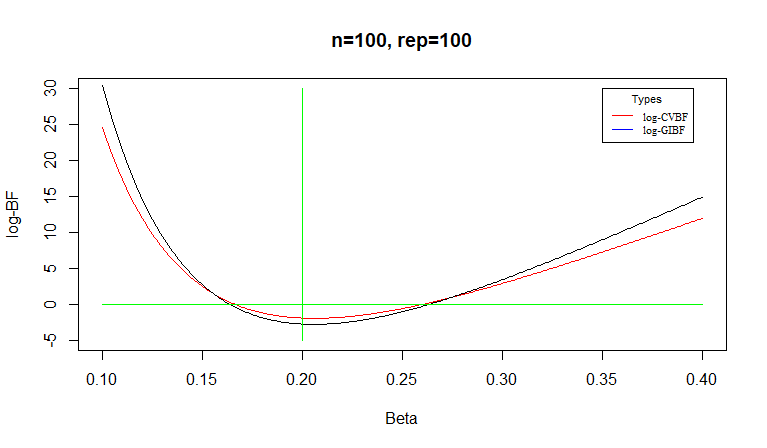}
    \caption{The Bridge Rule Works under Different Values of the Parameter}
    \label{Figure 6}
    \medskip
    \small
    In the figure, we generate the data of the size of 100. We Calculate the log of CVBF and the log of GIBF with 100 replicas when varying the parameter values. The red line and black line refer to means of all computed log-CVBF and log-GIBF, respectively. 

\end{figure}

\section{Two-parameter in normal means problem}

With working out the one-parameter case, here we would like to look into a two-parameter normal means problem to check the consistency.

Suppose we have a hypothesis testing that the first half of data points have mean $\mu_{1}$ and second half of data points have mean $\mu_{2}$. The system can be expressed by \[Y=\begin{pmatrix} 1 & 0 \\ \vdots  & \vdots  \\ 1 & 0 \\ 0 & 1 \\ \vdots  & \vdots  \\ 0 & 1 \end{pmatrix}\begin{pmatrix} { \mu  }_{ 1 } \\ { \mu  }_{ 2 } \end{pmatrix}+{ \varepsilon  }_{ i },\] where ${ \varepsilon  }_{ i }\sim N(0,{ \sigma  }^{ 2 }I)$, $i=1,...,n$ and ${ \sigma  }^{ 2 }$ is known.

The hypothesis testing is 
    \[{ H }_{ 0 }:\quad { \mu  }_{ 1 }={ \mu  }_{ 2 }\quad vs\quad { H }_{ 1 }:\quad { \mu  }_{ 1 }\neq { \mu  }_{ 2 }\]
    
The priors for $\mu_{1}$ and $\mu_{2}$ are both identical uniform distributions. In this regard, the corrected IBF is going to be \[{ { BF }_{ 1,0 } }^{ cI }=\sqrt { \frac { em }{ n }  } exp\{ -\frac { 1 }{ 2 } [-{ { \tilde { { Z1 }_{ 1 } }  }^{ 2 }+ }{ \tilde { { Z2 }_{ 1 } }  }^{ 2 }-{ \tilde { { Z3 }_{ 1 } }  }^{ 2 }+{ \tilde { { Z4 }_{ 1 } }  }^{ 2 }-{ \tilde { { Z5 }_{ 1 } }  }^{ 2 }+{ \tilde { { Z6 }_{ 1 } }  }^{ 2 }]\},\]   where ${ \tilde { { Z1 }_{ 1 } }  }^{ 2 }$, ${ \tilde { { Z2 }_{ 1 } }  }^{ 2 }$,..., ${ \tilde { { Z6 }_{ 1 } }  }^{ 2 }$ are non-central Chi-squared distributions with 1 degree of freedom and non-central parameters $\frac { m{ ({ \mu  }_{ 1 }+{ \mu  }_{ 2 }) }^{ 2 } }{ 4{ \sigma  }^{ 2 } },$ $\frac { n{ ({ \mu  }_{ 1 }+{ \mu  }_{ 2 }) }^{ 2 } }{ 4{ \sigma  }^{ 2 } },$ $\frac { \frac { n }{ 2 } { { \mu  }_{ 1 } }^{ 2 } }{ { \sigma  }^{ 2 } },$ $\frac { \frac { m }{ 2 } { { \mu  }_{ 1 } }^{ 2 } }{ { \sigma  }^{ 2 } },$ $\frac { \frac { n }{ 2 } { { \mu  }_{ 2 } }^{ 2 } }{ { \sigma  }^{ 2 } }$ and $\frac { \frac { m }{ 2 } { { \mu  }_{ 2 } }^{ 2 } }{ { \sigma  }^{ 2 } }$, respectively. $m$ is the training sample size, here $m=2$.

CVBF is going to be \[{ { BF }_{ 1,0 } }^{ CV }=exp\{ -[\frac { (n-m) }{ 2m } { { { \tilde { { Z }_{ 1 } }  } }^{ 2 }+ }\frac { n }{ 4m } { { Z }_{ 2 } }^{ 2 }+\frac { n }{ 4m } { { Z }_{ 3 } }^{ 2 }-\frac { n }{ 4m } { { \tilde { { Z }_{ 4 } }  } }^{ 2 }-\frac { n }{ 4m } { { \tilde { { Z }_{ 5 } }  } }^{ 2 }]\}, \] where ${ { Z }_{ 2 } }^{ 2 }$ and ${ { Z }_{ 3 } }^{ 2 }$ are standard Chi-squared distributions and ${ \tilde { { Z }_{ 1 } }  }^{ 2 }$, ${ \tilde { { Z }_{ 4 } }  }^{ 2 }$, ${ \tilde { { Z }_{ 5 } }  }^{ 2 }$ are non-central Chi-squared distributions with 1 degree of freedom and non-central parameter $\frac { m{ ({ \mu  }_{ 1 }-{ \mu  }_{ 2 }) }^{ 2 } }{ 4{ \sigma  }^{ 2 } } $, $\frac { m(n-m){ ({ \mu  }_{ 1 }-{ \mu  }_{ 2 }) }^{ 2 } }{ 2n{ \sigma  }^{ 2 } } $ and $\frac { m(n-m){ ({ \mu  }_{ 1 }-{ \mu  }_{ 2 }) }^{ 2 } }{ 2n{ \sigma  }^{ 2 } } $, respectively.

The expectation of the log of corrected IBF (m should be set as 2) is going to be \[E(log{ { BF }_{ 1,0 } }^{ cI }|{ M }_{ 1 })=E(log{ { BF }_{ 1,0 } }^{ cGI }|{ M }_{ 1 })=\frac { 1 }{ 2 } log\frac {  em }{ n } +\frac { 1 }{ 2 } \frac { (n-m){ ({ \mu  }_{ 2 }-{ \mu  }_{ 1 }) }^{ 2 } }{ 4{ \sigma  }^{ 2 } }. \]

The expectation of the log of CVBF is \[E(log{ { BF }_{ 1,0 } }^{ CV }|{ M }_{ 1 })=E(log{ { AvgBF }_{ 1,0 } }^{ CV }|{ M }_{ 1 })=-\frac { n-m }{ 2m } +\frac { 1 }{ 2 } \frac { (n-m){ ({ \mu  }_{ 2 }-{ \mu  }_{ 1 }) }^{ 2 } }{ 4{ \sigma  }^{ 2 } } ,\] when we apply the bridge rule, it becomes \[E(log{ { BF }_{ 1,0 } }^{ CV }|{ M }_{ 1 })=E(log{ { AvgBF }_{ 1,0 } }^{ CV }|{ M }_{ 1 })=-log(\frac { n }{ 2 } )+\frac { 1 }{ 2 } +\frac { 1 }{ 2 } \frac { (n-\frac { n }{ log(\frac { n }{ 2 } ) } ){ ({ \mu  }_{ 2 }-{ \mu  }_{ 1 }) }^{ 2 } }{ 4{ \sigma  }^{ 2 } } .\]

We introduce a updated rule for training sample sizes of CVBF, which is \[\frac { N }{ log(\frac { N }{ K } ) },\] where $K$ is the number of parameters. In this case, the rule is $\frac { N }{ log(\frac { N }{ 2 } ) }$, which forces IBF and CVBF to be equivalent in expectations when the null hypothesis model is valid. IBF and CVBF are both going to $-\infty$ at a rate of $n$ when $n$ goes to $\infty$ under the null model; they are going to $\infty$ at a rate of $exp(n)$ when $n$ goes to $\infty$ under the alternative. Hence, we have verified that they are consistent in this setting.

\section{Unknown-variance in normal means}

We here analyze a hypothesis testing with a null hypothesis following a normal distribution $N(0,\sigma^{2})$ and the alternative following a normal distribution $N(\theta,\sigma^{2})$, where $\theta$ is unknown and different from $0$, and $\sigma^{2}$ is unknown. Notice that this testing is quite different from Section 3.1. For convenience and simplicity, we use $1/\sigma$ as a prior distribution to the null model and $1/\sigma^{2}$, that is, a modified Jefferys prior in \cite{berger1996intrinsic}, as a prior distribution to an alternative model for computing GIBF. 

\subsection{Expressions of GIBF and CVBF}

In this case, we use the number of parameters as a training sample size, which is 2, the number of parameters. Hence, IBF is \[{ { BF }_{ 1,0 } }^{ I }({ { x }_{ i } }^{ (l) },{ { x }_{ j } }^{ (l) })=\sqrt { \frac { 2\pi  }{ n }  } { (1+\frac { n{ { \bar { x }  } }^{ 2 } }{ { { s }^{ (l) } }^{ 2 } } ) }^{ \frac { n }{ 2 }  }(\frac { { ({ { x }_{ i } }^{ (l) }-{ { x }_{ j } }^{ (l) }) }^{ 2 } }{ 2\sqrt { \pi  } ({ { { x }_{ i } }^{ (l) } }^{ 2 }+{ { { x }_{ j } }^{ (l) } }^{ 2 }) } ),\] where ${{s}^{(l)}}^{2}=\sum _{ i=1 }^{ n }{ { ({{ x }_{ i }}^{(l)}-{\bar { x }} ) }^{ 2 } },$ ${x_{i}}^{(l)}$ and ${x_{j}}^{(l)}$ are two data points, which are training samples. 

Furthermore, after taking geometric mean, GIBF is going to be \[{{ { BF }_{ 1,0 } }^{ GI }}({{ { x }_{ i } }^{ (l) }},{{ { x }_{ j } }^{ (l) }})=\sqrt { \frac { 2\pi  }{ n }  } { { (1+\frac { {n { \bar { x }  }^{{ (l) } ^{ 2 }}} }{ { { s }^{ (l) } }^{ 2 } } ) } }^{ \frac { n }{ 2 }  }{\prod _{ l=1 }^{ L }}{{ { (\frac { { ({{ { x }_{ i } }^{ (l) }}-{{ { x }_{ j } }^{ (l) }}) }^{ 2 } }{ 2\sqrt { \pi  } ({ { { x }_{ i } }^{ (l) } }^{ 2 }+{{ { { x }_{ j } }^{ (l) } }^{ 2 }}) } ) } } }^{ \frac { 1 }{ L }  } \]

The expression of log of IBF is \[log{ { BF }_{ 1,0 } }^{ I }=\frac { 1 }{ 2 } log(\frac { 2\pi  }{ n } )+\frac { n }{ 2 } log({ \chi  }^{ 2 }(1,\frac { { \theta  }^{ 2 } }{ { \sigma  }^{ 2 } } ))-\frac { n }{ 2 } log({ \chi  }^{ 2 }(n-1))+log(2{ \chi  }^{ 2 }(1))-log({ \chi  }^{ 2 }(2,\frac { { 2\theta  }^{ 2 } }{ { \sigma  }^{ 2 } } )),\] where ${ \chi  }^{ 2 }(1,\frac { { \theta  }^{ 2 } }{ { \sigma  }^{ 2 } } )$, ${ \chi  }^{ 2 }(n-1)$, ${ \chi  }^{ 2 }(1)$ and ${ \chi  }^{ 2 }(2,\frac { { 2\theta  }^{ 2 } }{ { \sigma  }^{ 2 } } )$ are Chi-squared distributions with $1$ degree of freedom and non-centrality $\frac { { \theta  }^{ 2 } }{ { \sigma  }^{ 2 } } $, $1$ degree of freedom and non-centrality $0$, $n-1$ degree of freedom and non-centrality $0$, and $2$ degree of freedom and non-centrality $\frac { 2{ \theta  }^{ 2 } }{ { \sigma  }^{ 2 } } $, respectively.

The expectation can only be evaluated as an infinite series (see \cite{berger1996justification}), but numerical solutions are straightforward. As \cite{berger1996intrinsic} claimed, one can simulate the expectation with parameters using the MLE of the original data. 

Here we do not correct IBF since the expectation is an infinite series. Fortunately, the corrected factor is just a constant; in this regard, we can analyze the consistency with or without the correction because of the minimal error.

Similarly, CVBF will be encountering difficulties. As other cases, one uses MLE to estimate parameters. In this case, there are two parameters, which are the mean and the variance. It is simply to compute the MLEs, $\hat { \theta  } =\bar { x } $ and $\hat { { \sigma  }^{ 2 } } =\frac { \sum _{ i=1 }^{ n }{ { ({ x }_{ i }-\bar { x } ) }^{ 2 } }  }{ n } $.

After partitioning data into a training set and a validation set, CVBF becomes\[\\ { { BF }_{ 1,0 } }^{ CV }={ (\frac { \sum _{ i=1 }^{ m }{ { { x }_{ i } }^{{ (l) }^{ 2 }} }  }{ \sum _{ i=1 }^{ m }{ { ({ x }_{ i }^{ (l) }-{ \bar { x }  }_{ m }^{ (l) }) }^{ 2 } }  } ) }^{ \frac { n-m }{ 2 }  }exp\{ -\frac { m }{ 2 } [\frac { \sum _{ i=m+1 }^{ n }{ { ({ x }_{ i }^{ (l) }-{ \bar { x }  }_{ m }^{ (l) }) }^{ 2 } }  }{ \sum _{ i=1 }^{ m }{ { ({ x }_{ i }^{ (l) }-{ \bar { x }  }_{ m }^{ (l) }) }^{ 2 } }  } -\frac { \sum _{ i=m+1 }^{ n }{ { { x }_{ i } }^{{ (l) }^{ 2 }} }  }{ \sum _{ i=1 }^{ m }{ { { x }_{ i } }^{{ (l) }^{ 2 }} }  } ]\},  \] where ${ \bar { x }  }_{ m }^{ (l) }$ is the mean of the training set.

Then the log of CVBF is \[log{ { BF }_{ 1,0 } }^{ CV }=\frac { n-m }{ 2 } [log({ \chi  }^{ 2 }(m,\frac { m{ \theta  }^{ 2 } }{ { \sigma  }^{ 2 } } ))-log({ \chi  }^{ 2 }(m-1))]-\frac { m }{ 2 } [\frac { \frac { m+1 }{ m } { \chi  }^{ 2 }(n-m,\frac { (n-m)m{ \theta  }^{ 2 } }{ { \sigma  }^{ 2 }(m+1) } ) }{ { \chi  }^{ 2 }(m-1) } -\frac { { \chi  }^{ 2 }(n-m,\frac { (n-m){ \theta  }^{ 2 } }{ { \sigma  }^{ 2 } } ) }{ { \chi  }^{ 2 }(m,\frac { m{ \theta  }^{ 2 } }{ { \sigma  }^{ 2 } } ) } ],\] where ${ \chi  }^{ 2 }$ is a random variable with a Chi-squared distribution.

However, the expectation will be an infinite series, which includes a confluent hypergeometric function of the first kind. 

Luckily, we can still work out the expectations of IBF and CVBF under the null hypothesis model. When the null is true, the expectation of log of IBF becomes \[E(log{ { BF }_{ 1,0 } }^{ I }|{ M }_{ 0 })=E(log{ { BF }_{ 1,0 } }^{ GI }|{ M }_{ 0 })=log(\sqrt { \frac { 1 }{ 8n }  } )+\frac { n }{ 2 } [log(1+\frac { 1 }{ n-3 } )-\frac { 1 }{ (n-2)(n-3) } ],\] and the expectation of log of CVBF becomes \[E(log{ { BF }_{ 1,0 } }^{ CV }|{ M }_{ 0 })=\frac { n-m }{ 2 } [log(1+\frac { 1 }{ m-3 } )-\frac { 1 }{ (m-2)(m-3) } ]-\frac { (m+1)(n-m) }{ 2(m-3) } +\frac { m(n-m) }{ 2(m-2) }, \] in which we use the rule $\frac { N }{ log(\frac { N }{ 2 } ) } $ so that they have consistency when $n$ is large. In the next section, we will be analyzing the consistency under the alternative by simulations. 

\subsection{Simulations}

Since the null model follows a normal distribution $N(0,\sigma^{2})$ and the alternative follows a normal distribution $N(\theta,\sigma^{2})$, we here suppose the sampling model is exactly the null model. The values of the log of Bayes Factors should be smaller than $0$. Hence, we generate the data from the null hypothesis, in which we use a normal distribution of $N(0,1)$. As we can observe in Figure 7.
\begin{figure}
    \centering
    \includegraphics[scale=0.53]{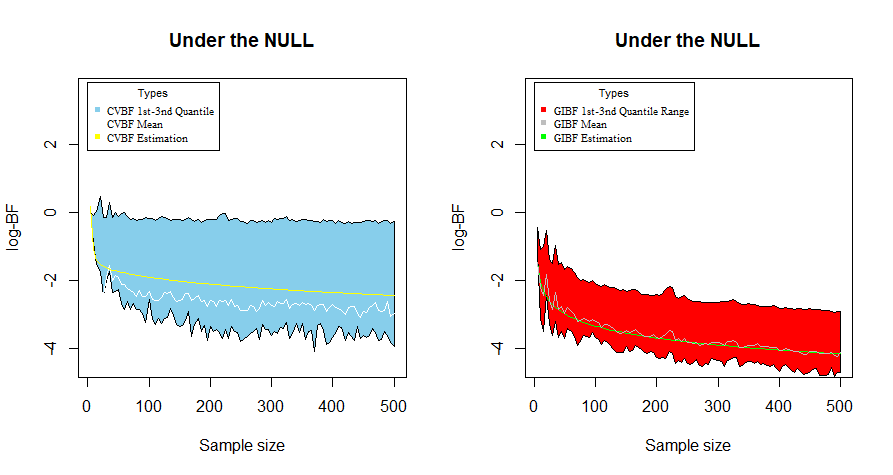}
    \caption{Consistency under the Null in Unknown $\sigma$ Normal Case}
    \medskip
    \small
    \label{Figure 7}
    In this figure we fix the training sets of GIBF as 2, and automatically, training sample sizes of CVBF become ${ m }_{ CV }=\frac { N }{ log(\frac { N }{ 2 } ) }  $. Moreover, we vary the sample sizes from 5 to 500 with a spacing of 5. The red area is the range between the first quantile and the third quantile on 1000 simulations of the log of GIBF; The sky blue area is the range between the first quantile and the third quantile on 1000 simulations of the log of CVBF. In the meantime, the white line and grey line are means on 1000 simulations of the log of CVBF and GIBF, respectively. The yellow line is the theoretical result of the expectation for CVBF; the green line is the expectation for GIBF.
\end{figure}

On the other hand, we would like to test the behaviors of the log of Bayes Factors when the alternative is the sampling model. We assume the true model is a normal distribution of $N(1,1)$. Then we generate the data from the model and then compute the Bayes factors. As we illustrate in Figure 8.
\begin{figure}
    \centering
    \includegraphics[scale=0.5]{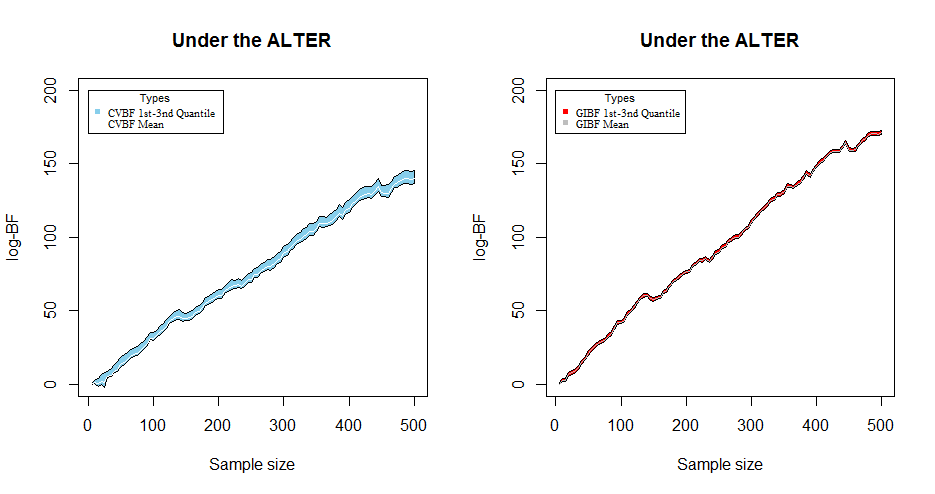}
    \caption{Consistency under the Alternative in Unknown $\sigma$ Normal Case}
    \label{Figure 8}
    \medskip
    \small
    In this figure we fix the training sets of GIBF as 2, and automatically, training sample sizes of CVBF become ${ m }_{ CV }=\frac { N }{ log(\frac { N }{ 2 } ) }  $. Moreover, we vary the sample sizes from 5 to 500 with a spacing of 5. The red area is the range between the first quantile and the third quantile on 1000 simulations of the log of GIBF; The sky blue area is the range between the first quantile and the third quantile on 1000 simulations of the log of CVBF. In the meantime, the white line and grey line are means on 1000 simulations of the log of CVBF and GIBF, respectively.

\end{figure}

\subsection{Summary}

The CVBF has the consistency with GIBF under both the null model and the alternative model based on the simulations. Expectations of CVBF and GIBF coincide with the simulations.  Although CVBF still has a slight gap with GIBF, most importantly, they have the same magnitudes. In Bayesian model selections, we would like to choose a better model over other models. Therefore, the crucial thing is that we can select the correct model instead of obtaining an exact value of a Bayes factor. The only sacrifice of using CVBF is the large variability.

\section{A Real Data Example (Civil Engineering Data)}

In this section, we would like to simulate the same data as the one in the paper \cite{hart2019prior} proposed, which are extracted from UC Irvine Machine Learning Repository 1030 determinations of $\, Y =\,$ concrete strength under a variety of different settings for the following eight design variables. $\, X_{1} =\,$ kg cement, $\, X_{2} =\,$ kg blast furnace slag, $\, X_{3} =\,$ kg fly ash, $\, X_{4} =\,$ kg water, $\, X_{5} =\,$ kg superplasticizer, $\, X_{6} =\,$ kg coarse aggregate, $\, X_{7} =\,$ fine aggregate and $\, X_{8} =\,$ age (in days).

Their null hypothesis model is going to be \[ Y\, =\, { \beta  }_{ 0 }+{ \beta  }_{ 1 }X_{ 1 }+{ \beta  }_{ 2 }{ X }_{ 2 }+{ \beta  }_{ 3 }{ X }_{ 3 }+{ \beta  }_{ 4 }{ X }_{ 4 }+{ \beta  }_{ 5 }{ X }_{ 5 }+{ \beta  }_{ 6 }{ X }_{ 6 }+{ \beta  }_{ 7 }{ X }_{ 7 }+{ \beta  }_{ 8 }{ X }_{ 8 }+{ \beta  }_{ 9 }\sqrt { { X }_{ 8 } } +{ \varepsilon  }_{ i },\] where $\, { \varepsilon  }_{ i }\sim N(0,exp({ a }_{ 0 }))$ and ${ a }_{ 0 }$ is a parameter. Hence, \[ { Y }_{ i }\sim N({ \beta  }_{ 0 }+{ \beta  }_{ 1 }X_{ 1 }+{ \beta  }_{ 2 }{ X }_{ 2 }+{ \beta  }_{ 3 }{ X }_{ 3 }+{ \beta  }_{ 4 }{ X }_{ 4 }+{ \beta  }_{ 5 }{ X }_{ 5 }+{ \beta  }_{ 6 }{ X }_{ 6 }+{ \beta  }_{ 7 }{ X }_{ 7 }+{ \beta  }_{ 8 }{ X }_{ 8 }++{ \beta  }_{ 9 }\sqrt { { X }_{ 8 } } ,exp({ a }_{ 0 })),\] therefore, the model is called a  homoscedastic model.

\begin{figure}
    \centering
    \includegraphics[scale=0.3]{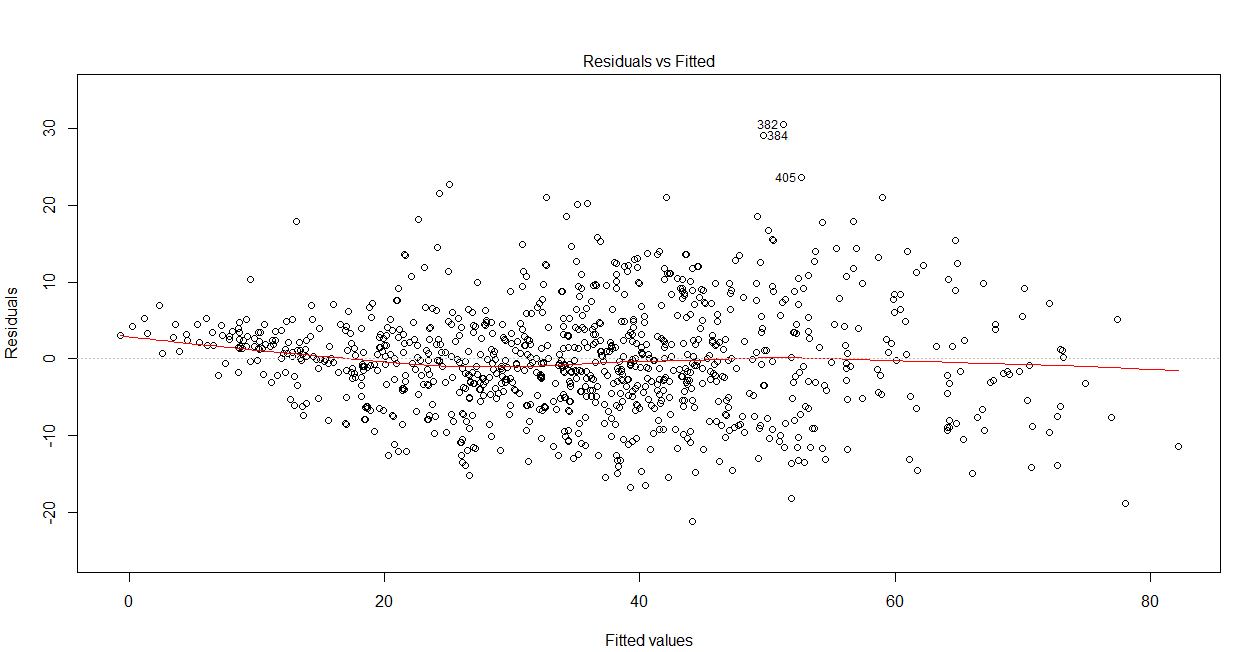}
    \caption{Residuals vs Fitted Values}
    \label{Figure 9}
    \medskip
    \small
    In this figure, the red line is the mean and the points are the coordinates of the residuals and means. 
\end{figure}

However, when we consider the null hypothesis as an accepted model, residuals from the fitted model are plotted against predicted values in Figure 9. The figure hints that the variance of error terms increases with the mean, which indicates the heteroscedasticity of the data. We wish to see if the model for this increase would be judged significantly better than the homoscedastic model by the use of Cross-validation Bayes factors.

In that regard, the alternative model is using the same model but has different variance for every error term, which is called a heteroscedastic model. And they define \[ { \varepsilon  }_{ i }\sim N(0,exp({ a }_{ 0 }+{ a }_{ 1 }(Z))),\] where $Z={ \beta  }_{ 0 }+{ \beta  }_{ 1 }X_{ 1 }+{ \beta  }_{ 2 }{ X }_{ 2 }+{ \beta  }_{ 3 }{ X }_{ 3 }+{ \beta  }_{ 4 }{ X }_{ 4 }+{ \beta  }_{ 5 }{ X }_{ 5 }+{ \beta  }_{ 6 }{ X }_{ 6 }+{ \beta  }_{ 7 }{ X }_{ 7 }+{ \beta  }_{ 8 }{ X }_{ 8 }+{ \beta  }_{ 9 }\sqrt { { X }_{ 8 } }$ and ${ a }_{ 1 }$ is another parameter. thus, \[ { Y }_{ i }\sim N(Z, exp({ a }_{ 0 }+{ a }_{ 1 }(Z ))).\]

Hence, the null hypothesis model and the alternative model have the same mean for each data. However, in the alternative model, every data point has a different variance, while the null model has the same variance for every data point. It is easy to be aware that if $\, a_{1}=0,\,$ the alternative model is exact the null model. First, we fix $\, { \beta  }_{ 0 },\, { \beta  }_{ 1 }\, ,{ \beta  }_{ 2 },...,\, { \beta  }_{ 9 }\,$ by doing the linear model optimizations. In null model, we will only have one parameter, say, $\,{ { a }_{ 0 } }^{ ' }.\,$ In order to obtain a  maximum likelihood estimator, we differentiate the likelihood and then equal it to zero for computing the optimal value of $\,{ { a }_{ 0 } }^{ ' },\,$ for which we evaluate in the likelihood such that we could obtain the maximum likelihood of the null model. As for the alternative model, there are two parameters in the model, which are $\,{ { a }_{ 0 } }\,$ and $\,{ { a }_{ 1 } }.\,$ After a similar process, we would attain the maximum likelihood of the alternative model. Finally, if we calculate the ratio of the maximum likelihood of the alternative and the maximum likelihood of the null model, CVBF with one simulation is attained.  

In the simulations, CVBFs were computed using seven choices for training sample size (denoted by $m$): 50, 100, 200, 300, 400, 500, and 600. For each of $m$, 200 random splits of data were considered (Different training sets are dependent on each other).

\begin{figure}
    \centering
    \includegraphics{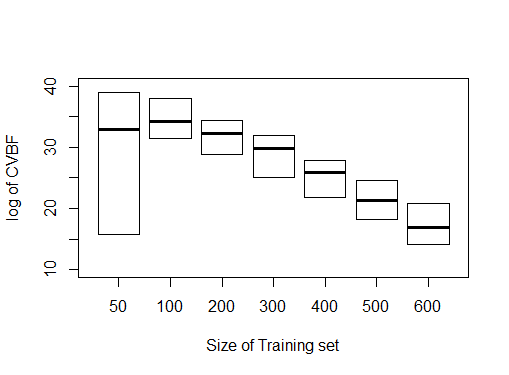}
    \caption{Boxplot of the Simulations}
    \label{Figure 10}
    \medskip
    \small
    In this figure, we present the boxplot of the log of CVBF. The boxes include the range from the first quantile to the third quantile, and the lines in the middle are the medians under different training set.
\end{figure}

In Figure 10, at each of m, the plot shows the median and the quantiles of 200 values of $log(CVBF_{1,0})$. As we know, $log(CVBF_{1,0}) \,>\, 0$ indicates that an alternative model is better than the null model. Hence, the alternative model is much favorable as the values of $log(CVBF_{1,0})$ are much higher than $0$ at each $m$.

On the other side, the range between $100$ and $200$ seems to be an optimal range of training sample sizes. When m is small, roughly, lower than 100, the variance is large; when we enlarge the training sample size, the reduction of variance is obvious. What is more, Bridge Rule here is going to be ${ m }_{ CV }=\frac { n }{ log(\frac { n }{ 2 } ) }=165 ,$ since the number of parameters in this case is 2. The value of $165$ is in the range of the optimal range, which suggests that our Rule performs reasonably.

\section{Conclusions}

In the paper, we analyze a new correspondence of real Bayes Factors with the recently proposed Cross-Validation Bayes Factors. Several important examples, including normal means problems with a known or unknown variance,  an exponential model, and a more complex regression example. We propose a Bridge Rule for choosing training sample sizes for CVBF, that is, \[{ m }_{ CV }=\frac { N }{ log(\frac { N }{ K } ) }, \] where $N$ is the sample size, and $K$ is the number of parameters. With this Rule, CVBF is broadly consistent with the Geometric Mean of Intrinsic Bayes Factors (GIBF). According to the criterion of GIBF, the optimal training set size of GIBF is the number of parameters. $K$ inside of the logarithm acts as a function to offset the effects that arise from the corrections of GIBF.

The performance of CVBF is broadly similar to the one of GIBF under simulations and analytical calculations. Both methods have their advantages and shortcomings. On the one hand, GIBF has better stability, but its computations might be non-trivial; on the other hand, CVBF is straightforward because the implicit priors are automatically assessed by the Rule, and there is no need to integration to calculate the marginal density of the models, which could save us time and effort. However, its more substantial variability is a shortcoming, sometimes it may cause a bit of trouble. Fortunately, this problem can be alleviated by Trimming methods.

After assigning the Bridge Rule, CVBF becomes an attractive method, actually, a Bayesian approach, since it tries to imitate GIBF. It has an implicit prior, which makes CVBF Bayesian, in the sense that it behaves approximately as a Bayes Factor with a proper objective prior. In contrast, the disadvantage is its large variability that one may encounter. Besides, we aim to choose a model instead of computing an exact Bayes factor, and when the evidence is overwhelming, it is not crucial to approximate the GIBF tightly. Furthermore, CVBF is trying to catch up with GIBF with the Rule by discarding the prior distributions and integration, which significantly reduces time consumption.

It can be argued that the computation of constant $c$ that corrects the Geometric Intrinsic Prior may be hard to compute in practical problems. We put forward the Bridge Rule, obtained in Normal and Exponential problems, as an approximation for general use since it should lead to consistency under both models. It does not affect the rate of convergence. In other words, even though we have obtained the constant $c$ exactly in examples in this article, we propose it as a general rule, even when the constant $c$ is challenging to obtain. 

\section{Open Problem}

GIBF, with a minimal training set, has an excellent approximation to Bayesian Information Criterion (BIC). What as to the relationship between CVBF and BIC? We will analyze these two matters using normal distributions with known variance and unknown variance in this section. 

\subsection{Normal means problem with known variance}
For the normal mean $(\theta)$ problem \[H_{0}:\theta=\theta_{0} \quad vs \quad H_{1}: \theta\neq\theta_{0}\] with known variance, the expectation of log of corrected Geometric Intrinsic Bayes factor is \[E(logBF^{CGI}|M_{1})=\frac{1}{2}log\frac{e}{n}+\frac{1}{2}\frac{(n-1)(\theta-\theta_{0})^{2}}{{\sigma}^{2}}.\]
And expectation of log-CVBF is \[ E(log{ { BF }_{ 1,0 } }^{ CV }|{ M }_{ 1 })=E(log{ { AvgBF }_{ 1,0 } }^{ CV }|{ M }_{ 1 })=-\frac { 1 }{ 2 } (\frac { n }{ m } -1-\frac { (n-m){ (\theta -{ \theta  }_{ 0 }) }^{ 2 } }{ { \sigma  }^{ 2 } } ),\]where $m$ is the training sample size for CVBF.

On the other hand, the formula for BIC is given by \cite{berger2001objective} \[logB^{S}_{1,0}=-\frac{(k_{1}-k_{0})}{2}log(n)+log(L_{1}/L_{0}).\]
Applying this formula, we have the resulting $logBIC_{1,0}$ \[E(logB^{S}_{1,0}|M_{1})=-\frac{1}{2}log(n)+\frac{n}{2\sigma^{2}}(\theta-\theta_{0})^{2}.\]
Therefore, the ratio \[\frac{E(logBF^{CGI}|M_{1})}{E(logB^{S}_{1,0}|M_{1})}\rightarrow 1 \]as $n$ goes to infinity, which suggests that, under both null model and alternative, GIBF with a training sample size of 1 is in the convergence of means with BIC in this case. This section reproduces the result.

Furthermore, \[\frac{E(logBF^{CV}|M_{1})}{E(logBF^{cGI}|M_{1})}\rightarrow 1 \] by Section 3.3, which implies that \[\frac{E(logBF^{CV}|M_{1})}{E(logB^{S}_{1,0}|M_{1})}\rightarrow 1 ,\] which means CVBF with the bridge rule is in convergence of means with BIC as n goes to infinity. 

Other important contents are Prior-based Bayesian information criterion (PBIC) and another version of Prior-based Bayesian information criterion, we call PBIC*, which are introduced by \cite{bayarri2019prior}. Expectation of log of PBIC can be expressed as \[E(logPBIC_{1,0})=\frac{n(\theta-\theta_{0})^2}{2\sigma^{2}}-\frac{1}{2}log(1+n)+log(\frac{1-\exp{(-\frac{\theta^{2}}{1+n})}}{\sqrt{2}\frac{\theta^{2}}{1+n}}).\] And the expectation of log of PBIC* is \[E(logPBIC^{*})=-\frac{1}{2}log(1+n)+\frac{n(\theta-\theta_{0})^{2}}{2\sigma^{2}}+log(\frac{1-exp(-c_{i})}{\sqrt{2v_{i}c_{i}}},\]where $v_{i}=\frac{\theta^{2}}{1+n}$ and $c_{i}=min\{v_{i},1.3\}.$

Figure 11 observes the expectations for different methods when the alternative model is correct.
Under the alternative ($\theta_{0}=1, \theta=2$), PBIC, BIC, PBIC*, and GIBF are almost overlapping at small perturbation. 

\begin{figure}
    \centering
    \includegraphics[scale=0.7]{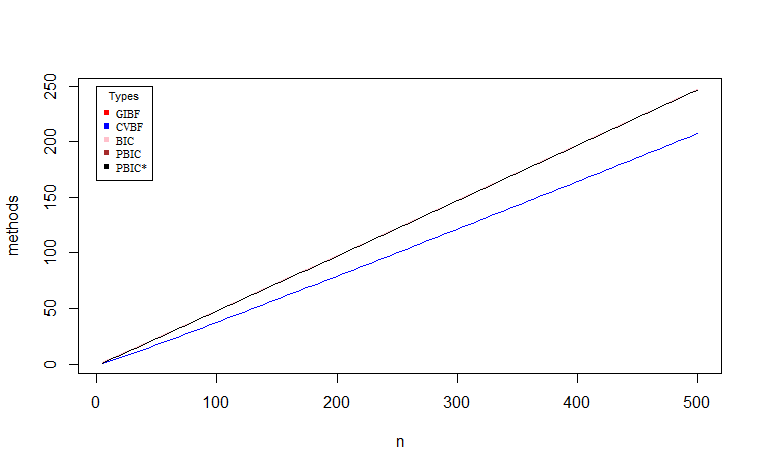}
    \caption{All methods under alternative model}
    \medskip
    \small
    In the graph, except for the blue line, all lines are overlapping or only being slightly different from others.
    \label{Figure 11}
\end{figure}

\subsubsection{Simulation on the normal distribution with known variance}

The hypothesis we suppose is that $H_{0}:$ $\theta=\theta_{0}=0$ versus $H_{1}:$ $\theta\neq\theta_{0}=0.$ we generate the data from $N(0,1)$ when null model is correct and generate the data from $N(1,1)$ when the alternative is true. Here we are listing the expressions of all methods (CVBF, GIBF, BIC, PBIC, PBIC*, etc.).

Another method is an asymptotic approximation of the Fractional Bayes factor (FBF), which is an improvement over BIC. The FBF approximation introduced in \cite{pericchi2005model} can be expressed as \[{ { B }_{ 1,0 } }^{ F }\approx \frac { { { f }_{ 1 } }^{ 1-b }(y|\theta ,{ \sigma  }^{ 2 }) }{ { { f }_{ 0 } }^{ 1-b }(y|{ \theta  }_{ 0 },{ \sigma  }^{ 2 }) } { b }^{ ({ k }_{ 1 }-{ k }_{ 0 })/2 },\] where $b=m/n$ and $k_{0}$ and $k_{1}$ are the number of parameters under model 0 and model 1, respectively. The log of this approximation will be \[log{ { B }_{ 1,0 } }^{ F }\approx (1-b)log[\frac { { { f }_{ 1 } }(y|\theta ,{ \sigma  }^{ 2 }) }{ { { f }_{ 0 } }(y|{ \theta  }_{ 0 },{ \sigma  }^{ 2 }) } ]+\frac { { k }_{ 1 }-{ k }_{ 0 } }{ 2 } logb.\]

Based on previous observations, log-CVBF is \[ logB_{1,0}^{CV}\, =\,  -\frac { (n-m) }{ 2{ \sigma  }^{ 2 } } [{ { ({ { { \bar { { x } }  }_{ m } }^{ (l) } } }-{ { \bar { { x } }  }_{ n-m } }^{ (l) }) }^{ 2 }-{ { ({ { { \bar { { x } }  }_{ n-m } }^{ (l) } } }-{ { \theta  }_{ 0 } }) }^{ 2 }] ,  \] where ${ \bar { { x } } _{ m } }^{ (l) }$ is the mean of the training samples while ${{ \bar { { x } }  }_{ n-m }}^{(l)}$ is the mean of the validation set. Specifically, $m$ here is $n/log(n).$

log-Corrected GIBF can be expressed by \[{ { logB }_{ 1,0 } }^{ cGI }=\frac { 1 }{ 2 } -\frac { 1 }{ 2 } log(n)-\frac { 1 }{ 2{ \sigma  }^{ 2 } } [\frac { \sum _{ i=1 }^{ n }{ { ({ x }_{ i }-{ \theta  }_{ 0 }) }^{ 2 } }  }{ n } -n{ (\bar { { x } } _{ n }-{ \theta  }_{ 0 }) }^{ 2 }].  \]

log-BIC is \[{ logB_{ 1,0 } }^{ S }=-\frac { 1 }{ 2 } log(n)+\frac { n }{ 2{ \sigma  }^{ 2 } } ({ { \theta  }_{ 0 } }^{ 2 }-{ { \theta  } }^{ 2 }+2\bar { x } (\theta -{ \theta  }_{ 0 })).\]

log-PBIC is \[{ logB_{ 1,0 } }^{ PB }=-\frac { 1 }{ 2 } log(1+n)+\frac { n }{ 2{ \sigma  }^{ 2 } } ({ { \theta  }_{ 0 } }^{ 2 }-{ { \theta  } }^{ 2 }+2\bar { x } (\theta -{ \theta  }_{ 0 }))+log(\frac { 1-{ e }^{ -\frac { { \theta  }^{ 2 } }{ 1+n }  } }{ \sqrt { 2 } \frac { { \theta  }^{ 2 } }{ 1+n }  } ).\]

log-PBIC* is \[{ logB_{ 1,0 } }^{ PB* }=-\frac { 1 }{ 2 } log(1+n)+\frac { n }{ 2{ \sigma  }^{ 2 } } ({ { \theta  }_{ 0 } }^{ 2 }-{ { \theta  } }^{ 2 }+2\bar { x } (\theta -{ \theta  }_{ 0 }))+log(\frac { 1-{ e }^{ -min\{ { v }_{ i },1.3\}  } }{ \sqrt { 2{ v }_{ i }min\{ { v }_{ i },1.3\}  }  } ),\] where ${ v }_{ i }=\frac { { \theta  }^{ 2 } }{ 1+n } .$

log-FBF approximation is \[log{ { B }_{ 1,0 } }^{ F }=-\frac { 1 }{ 2 } log(n)+(1-\frac { 1 }{ n } )\frac { { n }({ { \theta  }_{ 0 } }^{ 2 }-{ { \theta  } }^{ 2 }+2\bar { x } (\theta -{ \theta  }_{ 0 })) }{ 2{ \sigma  }^{ 2 } } .\]

And moreover, Arithmetic Intrinsic Bayes Factor (AIBF) is simply the arithmetic average of Intrinsic Bayes Factors, that is ${ { B }_{ 1,0 } }^{ AI }=\sum _{  }^{  }{ { { B }_{ 1,0 } }^{ I } } ,$ or $log{ { B }_{ 1,0 } }^{ AI }=log\sum _{  }^{  }{ { { B }_{ 1,0 } }^{ I } } .$

\begin{figure}
    \centering
    \includegraphics[scale=0.7]{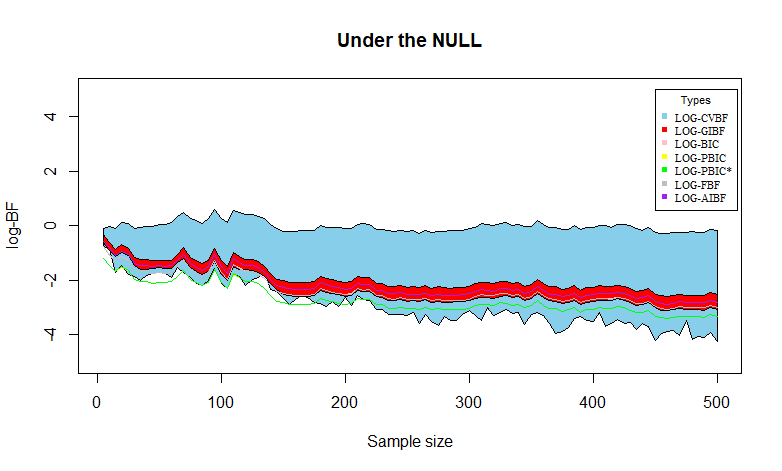}
    \caption{All types of methods under the null model}
    \medskip
    \small

    In this figure we fix the training sets of GIBF as 1, and automatically, training sample sizes of CVBF become ${ m }_{ CV }=\frac { N }{ log({ N } ) }  $. Moreover, we vary the sample sizes from 5 to 500 with a spacing of 5. The red area is the range between the first quantile and the third quantile on 1000 simulations of the log of GIBF; The sky blue area is the range between the first quantile and the third quantile on 1000 simulations of the log of CVBF. The pink line, yellow line, green line, grey line, and purple line are log-BIC, log-PBIC, log-PBIC*, log-FBF, log-AIBF, respectively. The grey line is the log of the Fractional Bayes factor. log-BIC is close to log-FBF, and log-PBIC is close to log-PBIC*.
    \label{Figure 12}
\end{figure}

\begin{figure}
    \centering
    \includegraphics[scale=0.7]{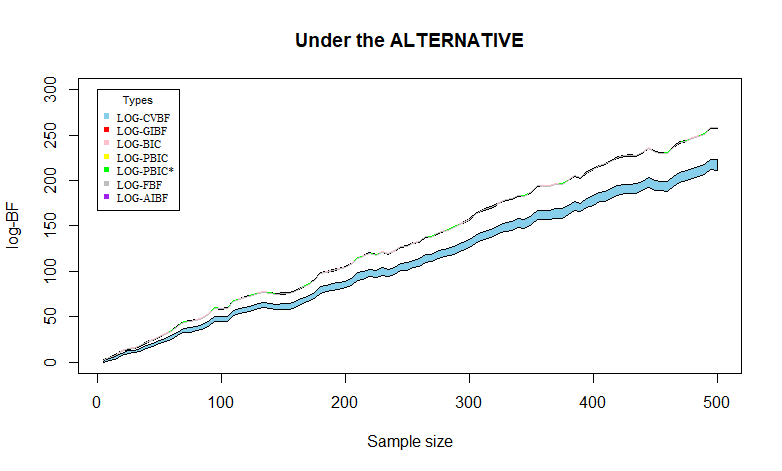}
    \caption{All types of methods under the alternative model}
        In this figure we fix the training sets of GIBF as 1, and automatically, training sample sizes of CVBF become ${ m }_{ CV }=\frac { N }{ log({ N } ) }  $. Moreover, we vary the sample sizes from 5 to 500 with a spacing of 5. The red area is the range between the first quantile and the third quantile on 1000 simulations of the log of GIBF; The sky blue area is the range between the first quantile and the third quantile on 1000 simulations of the log of CVBF. The pink line, yellow line, green line, grey line, and purple line are log-BIC, log-PBIC, log-PBIC*, log-FBF, log-AIBF, respectively. log-BIC, log-PBIC, log-PBIC*, log-FBF, and log-AIBF are almost overlapping. 

    \label{Figure 13}
\end{figure}

In the case, under both the null and the alternative, GIBF, BIC, PBIC, PBIC*, FBF approximation, and AIBF are very similar; see Figures 12 and 13.

\subsection{Normal means problem with unknown variance}

As stated in Section 6, we here analyze a hypothesis testing with a null hypothesis following a normal distribution $N(0,\sigma^{2})$ and the alternative following a normal distribution $N(\theta,\sigma^{2})$, where $\theta$ is unknown and different from $0$, and $\sigma^{2}$ is unknown. We use $1/\sigma$ as a prior distribution to the null model and $1/\sigma^{2}$ as a prior distribution to an alternative model for computing GIBF. 

In this case, we use the number of parameters as a training sample size, which is 2. Hence, IBF is \[{ { BF }_{ 1,0 } }^{ I }({ { x }_{ i } }^{ (l) },{ { x }_{ j } }^{ (l) })=\sqrt { \frac { 2\pi  }{ n }  } { (1+\frac { n{ { \bar { x }  } }^{ 2 } }{ { { s }^{ (l) } }^{ 2 } } ) }^{ \frac { n }{ 2 }  }(\frac { { ({ { x }_{ i } }^{ (l) }-{ { x }_{ j } }^{ (l) }) }^{ 2 } }{ 2\sqrt { \pi  } ({ { { x }_{ i } }^{ (l) } }^{ 2 }+{ { { x }_{ j } }^{ (l) } }^{ 2 }) } ),\] where ${{s}^{(l)}}^{2}=\sum _{ i=1 }^{ n }{ { ({{ x }_{ i }}^{(l)}-{\bar { x }} ) }^{ 2 } },$ ${x_{i}}^{(l)}$ and ${x_{j}}^{(l)}$ are two data points, which are training samples. 

After partitioning data into a training set and a validation set, CVBF becomes\[ { { BF }_{ 1,0 } }^{ CV }=\\\\{ (\frac { \sum _{ i=1 }^{ m }{ { { x }_{ i } }^{{ (l) }^{ 2 }} }  }{ \sum _{ i=1 }^{ m }{ { ({ x }_{ i }^{ (l) }-{ \bar { x }  }_{ m }^{ (l) }) }^{ 2 } }  } ) }^{ \frac { n-m }{ 2 }  }exp\{ -\frac { m }{ 2 } [\frac { \sum _{ i=m+1 }^{ n }{ { ({ x }_{ i }^{ (l) }-{ \bar { x }  }_{ m }^{ (l) }) }^{ 2 } }  }{ \sum _{ i=1 }^{ m }{ { ({ x }_{ i }^{ (l) }-{ \bar { x }  }_{ m }^{ (l) }) }^{ 2 } }  } -\frac { \sum _{ i=m+1 }^{ n }{ { { x }_{ i } }^{{ (l) }^{ 2 }} }  }{ \sum _{ i=1 }^{ m }{ { { x }_{ i } }^{{ (l) }^{ 2 }} }  } ]\},\]  where ${ \bar { x }  }_{ m }^{ (l) }$ is the mean of the training set.

The formula for the log of BIC is \[logB^{S}_{1,0}=-\frac{(k_{1}-k_{0})}{2}log(n)+log(L_{1}/L_{0}).\]
Applying this formula, we have the resulting $logBIC_{1,0}$ \[ logB^{S}_{1,0}|M_{1}=-\frac{1}{2}log(n)+\frac{n(2\bar{x} \theta-\theta^2)}{2\sigma^2}=-\frac{1}{2}log(n)+\frac{n(2\bar{x}^{2}-\bar{x}^{2})}{2\frac{\sum _{ i=1 }^{ n }{ { ({ x }_{ i }-\bar { x } ) }^{ 2 } }}{n}}=-\frac { 1 }{ 2 } log(n)+\frac { { n }^{ 2 }\bar { x } ^{ 2 } }{ 2\sum _{ i=1 }^{ n }{ { ({ x }_{ i }-\bar { x } ) }^{ 2 } }  } .\]

The log-PBIC in this case can be expressed by \[logPBIC_{1,0}=\frac { { n }^{ 2 }\bar { x } ^{ 2 } }{ 2\sum _{ i=1 }^{ n }{ { ({ x }_{ i }-\bar { x } ) }^{ 2 } }  }-\frac{1}{2}log(1+n)+log(\frac{1-e^{-\frac{\bar{x}^{2}}{1+n}}}{\sqrt{2}\frac{\bar{x}^{2}}{1+n}}).\] It is important to illustrate that PBIC is more favorable to the null hypothesis. 

Another version of PBIC will be in favor of the models which are in the middle of the models that PBIC and BIC favor. log-PBIC* is \[logPBIC^{*}_{1,0}=\frac { { n }^{ 2 }\bar { x } ^{ 2 } }{ 2\sum _{ i=1 }^{ n }{ { ({ x }_{ i }-\bar { x } ) }^{ 2 } }  }-\frac{1}{2}log(1+n)+log(\frac{1-e^{-min\{v_{i},1.3\}}}{\sqrt{2v_{i}min\{v_{i},1.3\}}}),\]where $v_{i}=-\frac{\bar{x}^{2}}{1+n}.$

In this example, $k_{0}=1$, $k_{1}=2$ and $m=2,$ which is the number of parameters. Computations yield $logFBF$ approximation \[log{ { B }_{ 1,0 } }^{ F }=-\frac { 1 }{ 2 } log(\frac { n }{ 2 } )+(1-\frac { 2 }{ n } )\frac { { n }^{ 2 }{ \bar { x }  }^{ 2 } }{ 2\sum _{ i=1 }^{ n }{ { ({ x }_{ i }-\bar { x } ) }^{ 2 } }  } .\]

On the other hand, after taking the arithmetic average of the IBF, and applying the logarithm, we will attain log-AIBF.

Since the null model follows a normal distribution $N(0,\sigma^{2})$ and the alternative follows a normal distribution $N(\theta,\sigma^{2})$, we here suppose the sampling model is precisely the null model. The values of the log of Bayes Factors should be smaller than $0$. Hence, we generate the data from the null hypothesis, in which we use a normal distribution of $N(0,1)$ as we can observe in Figure 14.

On the other hand, we would like to test the behaviors of the log of Bayes Factors when the alternative is the sampling model. At this point, we assume the correct model is a normal distribution of $N(1,1)$. Then we generate the data from the model and compute all kinds of Bayes factors as we illustrate in Figure 15.

\begin{figure}
    \centering
    \includegraphics[scale=0.7]{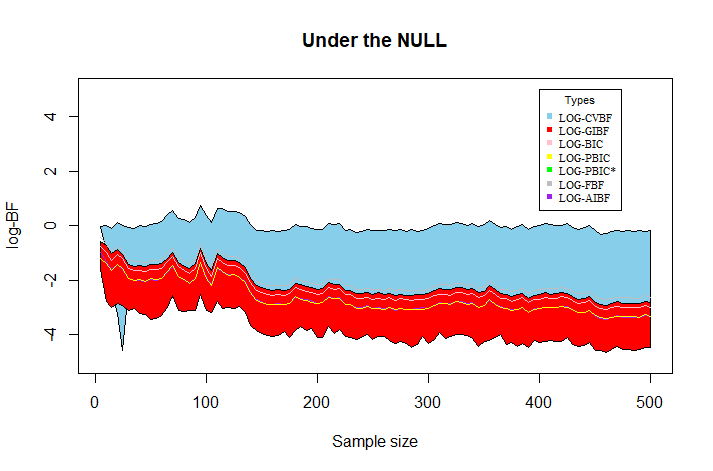}
    \caption{All types of methods under the null model}
    \medskip
    \small

    In this figure we fix the training sets of GIBF as 2, and automatically, training sample sizes of CVBF become ${ m }_{ CV }=\frac { N }{ log(\frac { N }{ 2 } ) }  $. Moreover, we vary the sample sizes from 5 to 500 with a spacing of 5. The red area is the range between the first quantile and the third quantile on 1000 simulations of the log of GIBF; The sky blue area is the range between the first quantile and the third quantile on 1000 simulations of the log of CVBF. The pink line, yellow line, green line, grey line, and purple line are log-BIC, log-PBIC, log-PBIC*, log-FBF, and log-AIBF, respectively. log-PBIC, log-PBIC*, and log-AIBF are almost overlapping. 
    \label{Figure 14}
\end{figure}

\begin{figure}
    \centering
    \includegraphics[scale=0.7]{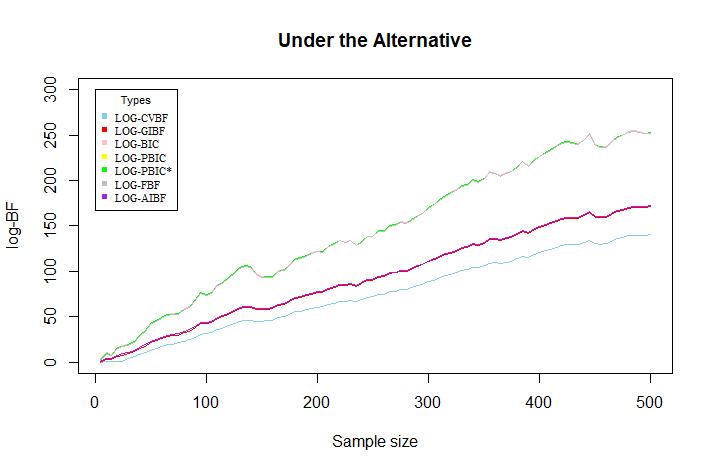}
    \caption{All types of methods under the alternative model}
        In this figure we fix the training sets of GIBF as 2, and automatically, training sample sizes of CVBF become ${ m }_{ CV }=\frac { N }{ log(\frac { N }{ 2 } ) }  $. Moreover, we vary the sample sizes from 5 to 500 with a spacing of 5. The red area is the range between the first quantile and the third quantile on 1000 simulations of the log of GIBF; The sky blue area is the range between the first quantile and the third quantile on 1000 simulations of the log of CVBF. The pink line, yellow line, green line, grey line, and purple line are log-BIC, log-PBIC, log-PBIC*, log-FBF, and log-AIBF, respectively. log-BIC, log-PBIC, log-PBIC*, and log-FBF are almost overlapping. 

    \label{Figure 15}
\end{figure}

CVBF is imitating well with GIBF. BIC is consistent with CVBF and GIBF under the null. However, it is more favorable to the alternative model than CVBF and GIBF under the alternative model assumption.

PBIC and PBIC* do not help a lot. BIC, PBIC, and PBIC* are almost overlapping with each other under the alternative. Under the alternative, they are overlapping since the difference between them is just around 1. For example, if we only compute the extra term that PBIC has compared with BIC. Let $n=500$ and generate the data from a normal distribution with mean one and variance one. Then the term $log(\frac{1-e^{-\frac{\bar{x}^{2}}{1+n}}}{\sqrt{2}\frac{\bar{x}^{2}}{1+n}})=-0.347.$ The difference is only -0.35, which is too small to reduce PBIC and PBIC*.

FBF approximation is also very similar to BIC. Moreover, AIBF has a similar trajectory to GIBF.

\subsection{Summary}
CVBF, GIBF, BIC, and other similar methods are quite close under the normal means problem with known variance. However, under the normal means problem with unknown variance, BIC, and variations from it, seems to be much favorable to the alternative when it is the sampling model. The results suggest that CVBF is a closer approximation to a real Bayes Factor like the GIBF, which would be a significant point in its favor. Overall, all the Bayes Factors and its approximations have the same tendency as we have proven for the convergence of means. As for why there exists an angle for the BIC and GIBF under the normal mean problem with unknown variance under the assumption of the alternative model, it is likely that both GIBF and CVBF are using cross-validation (in different ways) to obtain their priors, while BIC and related methods do not.

\bibliographystyle{unsrtnat}
\bibliography{sample}
\nocite{almodovar2012new}
\nocite{aitkin1991posterior}
\nocite{berger1996justification}

\nocite{berger2014bayes}
\nocite{bertsekas2002introduction}
\nocite{chen2010frontiers}
\nocite{fong2019marginal}
\nocite{stone1977asymptotic}
\nocite{jeffreys1998theory}

\nocite{kass1995bayes}
\nocite{pericchi2005model}

\nocite{yao2018using}

\end{document}